# X-ray Diffraction and Electrical Transport Imaging of Superconducting Superhydride (La,Y)H$_{10}$

A. H. Manayil Marathamkottil[1*], Kui Wang[2], Nilesh P. Salke[2], Muhtar Ahart[2], Alexander C. Mark[2], Ross Hrubiak[3], Stella Chariton[4], Dean Smith[3], Vitali B. Prakapenka[4], Maddury Somayazulu[3], Nenad Velisavljevic[3,5], Russell J. Hemley[1,2,6*]

[1]Department of Chemistry, University of Illinois Chicago, Chicago, IL 60607, USA

[2]Department of Physics, University of Illinois Chicago, Chicago, IL 60607, USA

[3]HPCAT, X-ray Science Division, Argonne National Laboratory, Lemont, IL 60439, USA

[4]Center for Advanced Radiation Sources, University of Chicago, Chicago, IL 60637, USA

[5]Physics Division, Lawrence Livermore National Laboratory, Livermore, CA 94550, USA

[6]Department of Earth and Environmental Sciences, University of Illinois Chicago, Chicago, IL 60607, USA

**Abstract**

We report the synthesis and characterization of (La$_{0.9}$Y$_{0.1}$)H$_{10}$ superhydrides exhibiting coexisting cubic $Fm\bar{3}m$ and hexagonal $P6_3/mmc$ clathrate phases observed over the pressure range from 168 GPa down to 136 GPa. Using synchrotron-based X-ray diffraction imaging (XDI) at the upgraded Advanced Photon Source (APS-U), we spatially resolved µm-scale distributions of these phases, revealing structural inhomogeneity across the sample. Four-probe DC resistance measurements confirmed superconductivity, with two distinct transitions: an onset at 244 K associated with the cubic phase and a second near 220 K linked to the hexagonal phase. Notably, resistance profiles collected from different current and voltage permutations showed variations in transition width and onset temperature that correlated with the spatial phase distribution mapped by XDI. These findings demonstrate a direct connection between local structural domains and superconducting behavior. Yttrium substitution is found to influence both the phase behavior and superconducting properties of LaH$_{10}$-type clathrate hydrides. More broadly, this study highlights the utility of spatially correlating structural and electrical transport measurements in materials exhibiting heterogeneity under pressure, including hydride superconductors.

# I. Introduction

Dense hydrides exhibiting very high temperature superconductivity are established as an extraordinary class of new materials of great fundamental and potential applied interest[1]. The remarkable quantum phenomena of these materials arise from the potential of dense hydrogen lattices to mimic atomic metallic hydrogen, which has long been predicted to exhibit room-temperature superconductivity under extreme pressures[2,3]. Advances in computational methods combined with experimental diamond anvil cell (DAC) techniques led to the discovery of rare-earth superhydrides with superconducting critical temperatures ($T_c$'s) approaching room temperature at megabar (>100 GPa) pressures[4–10]. In particular, $LaH_{10}$ was found to superconduct at temperatures up to 260 K at 188 GPa in the cubic $Fm\bar{3}m$ clathrate phase[8,9]. This structure features a three-dimensional hydrogen cage network surrounding La atoms, enabling the strong electron–phonon coupling origin of this very high $T_c$ superconductivity[4–7]. Notably, the near-room-temperature superconductivity was confirmed and has been independently reproduced by a growing number of groups[10–12].

Recent research has focused on strategies to expand the critical temperatures and stability of these materials by chemical substitution to form ternary and higher order phases[13–21]. For lanthanum superhydrides, theoretical predictions suggest that partial replacement of La with smaller elements like yttrium (Y) may chemically pre-compress the lattice, modify phonon spectra, and extend the stability field of the clathrate phase[13–15]. In particular, $(La,Y)H_{10}$ phases are predicted to retain the high symmetry of $LaH_{10}$ while gaining enhanced structural stability through Y incorporation[13–16]. Experimental work has shown $T_c$ values up to 253 K for $(La_{0.8}Y_{0.2})H_{10}$ at 183 GPa[22], yet a comprehensive understanding of the pressure–composition phase space remains incomplete[22,23].

Moreover, the coexistence of multiple structural phases, together with stress-strain gradients, in these and related samples further raises questions about how local structural heterogeneity affects superconducting behavior[10,21,22]. Traditional bulk measurements obscure this complexity, averaging over the entire sample volume and masking microscale phase variation or domain-level interactions[24,25]. To resolve such structure–property relationships, techniques capable of spatially correlating local structural and electronic properties are essential, especially under the extreme conditions of DAC experiments, where length scales of structure-property variations are in the μm to sub-μm range[26].

Synchrotron-based scanning X-ray diffraction microscopy (SXDM) has emerged as a powerful tool for such studies[26–29]. Enabled by the upgraded Advanced Photon Source (APS-U), SXDM provides enhanced spatial resolution and sensitivity to local symmetry variations, strain, and phase coexistence. When combined with the X-ray diffraction imaging (XDI) software for data visualization, this approach offers a direct window into heterogeneous structural states that are often hidden in conventional diffraction measurements[26,29]. Complementing this microstructural information, four-probe DC resistance measurements utilizing different current/voltage probe permutations allows for spatial sampling of superconducting behavior across a sample[30]. While originally designed to measure homogeneous sheets, the van der Pauw (VDP) algorithm may be modified to probe inhomogeneous materials and correlated to spatially resolved XDI measurements to capture broadened or multi-step transitions arising from structural inhomogeneity[30,31]. Such spatially sensitive transport techniques have rarely been applied to hydrides due to technical constraints in DACs.

In this work, we synthesized $(La_{0.9}Y_{0.1})H_{10}$ at high pressure and employed SXDM in combination with multi-channel electrical transport measurements. Using the ~1 μm focused X-ray beam at HPCAT-U, SXDM revealed μm-scale structural inhomogeneity, visualized through XDI, and correlated with local superconducting transitions captured via spatially resolved resistance measurements. This represents the first direct correlation between crystal structure and superconducting behavior in hydrides. Two distinct superconducting onsets near 244 K and 220 K were found to correspond to regions dominated by the cubic $Fm\overline{3}m$ and hexagonal $P6_3/mmc$ clathrate phases, respectively, enabling unambiguous assignment of $T_c$ values to specific structures. These results show that yttrium substitution stabilizes the coexistence of both phases down to 136 GPa, although it does not enhance the $T_c$ relative to pure $LaH_{10}$. Overall, our findings establish a clear, spatially resolved link between structural domains and superconducting behavior in hydrides.

## II. Results and Discussion

### A. Structural Characterization of (La,Y)H$_{10}$

Prior to laser heating, the $La_{0.9}Y_{0.1}$ alloy compressed with ammonia borane at high pressure exhibited a distorted-cubic *Fmmm* structure (Fig. 1 and Fig. S1, bottom panel)[32]. Following laser heating, the diffraction pattern changed significantly, with alloy peaks disappearing and new reflections emerging from hydrogen-rich phases (Fig. 1 and Fig. S1), confirming successful hydrogenation. Optical images of the sample before and after laser heating are provided in Fig. S2.

The first synthesis was performed at 172 GPa. Following laser heating, the pressure relaxed to 168 GPa. Synchrotron XRD at this pressure (Fig. S1) revealed the formation of two clathrate structures: cubic $Fm\bar{3}m$ and hexagonal $P6_3/mmc$, both of which have been previously reported in La–H and La–Y–H systems[10,22]. A second synthesis was conducted at 158 GPa using DAC #2 (Fig. 1); after laser heating, the pressure relaxed to 153 GPa. The same two phases were again observed, demonstrating reproducibility across different experimental runs.

Le Bail refinements of the 153 GPa XRD pattern yielded lattice parameters of $a = 5.15(1)$ Å for the $Fm\bar{3}m$ phase and $a = 3.71(1)$ Å, $c = 5.54(1)$ Å for the $P6_3/mmc$ phase, corresponding to unit cell volumes of 136.5(1) Å³ and 66.1(1) Å³, respectively (Fig. 1, top panel). These values are comparable to those reported for undoped $LaH_{10}$ at similar pressures[6], with only slight reductions in volume consistent with the expected lattice contraction from yttrium substitution[14,22]. Notably, no diffraction peaks from secondary $LaH_n$ or $YH_n$ phases were observed, indicating that ~10% Y remains within the solubility limit for forming a single-phase clathrate or a mixed-phase clathrate solid solution[13,22].

To assess the stability of the observed phases during decompression, additional XRD measurements were performed on both samples. In DAC #1, after initial synthesis at 172 GPa and characterization at 168 GPa, the sample was decompressed to 161 GPa. At this pressure, both the $Fm\bar{3}m$ and $P6_3/mmc$ phases remained clearly present (Fig. S3), confirming that the clathrate structures persist during the early stages of decompression. Similarly, the sample in DAC #2 was gradually decompressed from 153 GPa, and an XRD was carried out at 136 GPa. Electrical resistance measurements were performed in the intermediate pressure range. At 136 GPa, both the cubic and hexagonal phases were still observed (Fig. S4), with only minor peak shifts in the $Fm\bar{3}m$ phase across different regions of the sample (Fig. S5), likely resulting from uneven pressure distribution and lattice relaxation. This pressure lies near the known structural phase boundary of undoped $LaH_{10}$, where transitions to lower-symmetry structures such as $R\bar{3}m$ or $C2/m$ typically occur[6,12]. The persistence of both clathrate phases at 136 GPa suggests that Y substitution extends the structural stability of $LaH_{10}$-type phases to lower pressures than observed in the undoped system.

The consistent observation of both $Fm\overline{3}m$ and $P6_3/mmc$ phases across the entire pressure range studied (172–136 GPa) confirms that these structures coexist rather than represent distinct, pressure-stabilized phases. The presence of both phases in the same sample indicates that Y substitution does not strongly favor one structure over the other but instead promotes a mixed-phase state. This contrasts with the behavior of pure $LaH_{10}$, where the $Fm\overline{3}m$ phase typically dominates above ~150 GPa and transforms to lower-symmetry phases upon decompression[6,9–12]. The spatial variation and distribution of the cubic and hexagonal domains, further examined through diffraction imaging, highlight the complexity of phase coexistence near the clathrate stability boundary.

The pressure–volume (P–V) behavior of $(La_{0.9}Y_{0.1})H_{10}$, shown in Fig. S3, generally follows the trend reported for undoped $LaH_{10}$, but with modest phase-specific deviations. These variations suggest non-uniform compressibility effects and possible local strain contributions. Comparable trends were reported for $(La_{0.8}Y_{0.2})H_{10}$ synthesized at higher pressures[22], where clathrate structures remained stable without decomposition. Together, these results demonstrate that partial Y substitution preserves the hydrogen cage framework of $LaH_{10}$ and enables phase coexistence across a broader pressure range[22,23]. These structural findings establish a foundation for correlating local phase heterogeneity with superconducting behavior, as discussed in the next sections.

## B. Spatial Mapping of Structural Domains via SXDM and XDI

Spatial phase mapping at 153 GPa revealed μm-scale coexistence of $Fm\overline{3}m$ and $P6_3/mmc$ clathrate domains. Using SXDM at HPCAT-U, diffraction patterns were collected across a 30 μm × 30 μm region with 3 μm steps. XDI-based analysis identified phase-specific intensity distributions by integrating the first two Bragg reflections unique to each structure, producing two-dimensional maps of local phase domains.

Figure 2 presents the optical image of the laser-heated region at 153 GPa, 2D X-ray scan overview of the sample chamber, individual domain maps, and the composite spatial distribution of Pt (gray), $Fm\overline{3}m$ (red), and $P6_3/mmc$ (blue) $(La,Y)H_{10}$ phases. Electrode positions are annotated to support spatial correlation with electrical transport measurements. The $Fm\overline{3}m$ phase is localized in discrete clusters near Pt leads #2, #3, and #4, whereas the $P6_3/mmc$ phase forms a more continuous matrix, with prominent coverage between leads #1 and #4. Image analysis of the domain maps indicates that the red $Fm\overline{3}m$ phase covers approximately 42% of the mapped region, while the blue $P6_3/mmc$ phase accounts for 58%. This structural inhomogeneity likely reflects local variations in hydrogen content, laser heating profiles, or stress gradients across the sample[9,10,12]. Although such phase coexistence is common in multiphase hydride systems[10,22], the ability to directly image μm-scale domain structure at this resolution provides a valuable framework for linking local structural environments with superconducting behavior, as discussed in the following sections.

Additional raster scans were performed at 136 GPa after decompression. Two maps, a broader 50 μm × 50 μm grid and a focused 15 μm × 15 μm grid, were collected from the same central region and are shown in Fig. 2. In the larger scan, the $Fm\bar{3}m$ phase exhibits reduced intensity near Pt lead #4 compared to the 153 GPa map, while the $P6_3/mmc$ phase remains more uniformly distributed across the sample chamber. The composite map again shows dominance of the hexagonal phase between leads #1 and #4, and clustering of the cubic phase around leads #2 and #3. The smaller scan offers a higher-resolution view of the local phase distribution and confirms the persistence of structural heterogeneity upon decompression.

The use of micro-focused beam combined with SXDM enabled spatial mapping of phase-separated regions within the sample that may be challenging to resolve using conventional bulk XRD techniques. The spatial resolution in this study was chiefly governed by the ~1 μm beam size of APS-U and the small raster step size used during SXDM. Together, these parameters enabled fine spatial sampling across the sample chamber, allowing detection of μm-scale structural variations. Prior applications of XDI have demonstrated its effectiveness in visualizing structural gradients and preferred nucleation patterns in $FeH_x$[33], $H_3S$[34], $H_2$[35], and La–Y–Ce–H[36] systems. However, earlier studies often had overlapping grids or had limited phase assignment capability due to reduced flux or detector sensitivity. The ability to resolve discrete $Fm\bar{3}m$ and $P6_3/mmc$ domains across the sample provides unique insight into structural heterogeneity. These spatially resolved maps form the foundation for linking local phase composition with superconducting behavior, as discussed below.

## C. Superconductivity in Coexisting Phases

Four-probe DC resistance measurements were carried out on the $(La_{0.9}Y_{0.1})H_{10}$ sample in DAC #2 following structural characterization. As shown in Fig. 3A, resistance vs. temperature curves collected during warming cycles at four pressures between 153 GPa and 136 GPa consistently display two distinct superconducting transitions. At 153 GPa, the first resistance drop begins at $T_{c,onset}$ = 244 K, followed by a second transition near $T_{c,onset}$ = 220 K (Fig. S6). The total transition width of ΔT ≈ 28 K is unusually broad for DC transport measurement and is characteristic of phase coexistence and electronic heterogeneity[37].

To further investigate the origin of these transitions, the temperature dependence of eight partial resistance traces [$R_{ab,cd}(T)$] was collected using the standard VDP permutations, involving four voltage pairs and two current directions, as shown in Fig. 4. Each measurement configuration was overlaid onto the composite XDI phase map, enabling direct spatial correlation between structural domains and electronic behavior. Notably, configurations such as $R_{34,12}$, $R_{34,21}$, $R_{41,23}$, and $R_{41,32}$ exhibited a sharp superconducting drop near 240 K, with a narrow transition width of ΔT < 10 K. These measurements probed regions between electrodes #3 and #4 and between #4 and #1, with voltage recorded across electrodes #1 and #2 or #2 and #3. Based on the spatial maps shown in Fig. 2, these current–voltage pathways intersected domains where the $Fm\bar{3}m$ phase was concentrated, particularly near electrode #2. This spatial correlation supports the assignment of the higher-temperature superconducting transition to the cubic clathrate phase.

In contrast, other partial resistance configurations, such as $R_{12,34}$, $R_{12,43}$, $R_{23,41}$, and $R_{23,14}$ exhibited broader, two-step transitions, with onsets near 241 K and 218 K, respectively. These configurations passed through regions where both $Fm\bar{3}m$ and $P6_3/mmc$ phases were present, with a higher fraction of the hexagonal phase observed between electrodes #1 and #4. The lower-temperature transition was thus attributed to the $P6_3/mmc$ clathrate phase. This interpretation is consistent with previously reported superconducting transition temperatures in undoped and Y-substituted $LaH_{10}$ systems (Fig. 3B)[9,10,22]. Notably, both transitions occurred at lower onset temperatures than in pure $LaH_{10}$, where $T_c$ typically exceeds 250 K under similar pressures[9–11]. The observed $T_c$ suppression provides complementary evidence of successful Y incorporation and its influence on the electronic structure, particularly through added intermediate-frequency phonon modes[15,22]. Furthermore, the direct correlation between partial resistances and spatial phase distribution underscores the utility of structural mapping for interpreting superconducting transport behavior in mixed-phase systems.

To evaluate the pressure dependence of superconductivity in the coexisting clathrate phases of $(La_{0.9}Y_{0.1})H_{10}$, resistance measurements were performed during decompression from 153 GPa to 136 GPa (Fig. 3A). Across this pressure range, the resistance–temperature profiles consistently displayed two distinct superconducting transitions, indicative of phase coexistence. Additional support for superconductivity comes from current–voltage (*I–V*) measurements at 146 GPa and 136 GPa (Fig. S7), which exhibit nonlinear behavior below ~230 K at 146 GPa. Notably, the overall resistance behavior and partial resistance traces remained similar down to 142 GPa (Fig. S8). Consistent with measurements at higher pressure, all transition temperatures at each pressure during decompression occurred within experimental uncertainty across the eight configurations, confirming that the observed features are intrinsic to the sample. At 136 GPa, a marked departure from the previous $R_h$–T profiles was observed (Fig. 3). The two-step superconducting features became less distinct, and the transition width narrowed ($\Delta T \sim 20$ K), particularly in partial resistances such as $R_{12,34}$, $R_{12,43}$, $R_{23,41}$, and $R_{23,14}$ (Fig. S9). These current paths intersect regions where the spatial phase map (Fig. 2) indicated reduced $Fm\bar{3}m$ intensity near Pt lead #4, consistent with diminished cubic phase contributions at lower pressures.

Meanwhile, traces such as $R_{34,12}$, $R_{34,21}$, $R_{41,23}$, and $R_{41,32}$ still exhibited sharp resistance drops, with more pronounced secondary features. These traces traverse regions around Pt leads #2 and #3, where the $Fm\bar{3}m$ phase remained spatially concentrated even after decompression. The persistence of sharp transitions in these configurations suggests that residual cubic domains retain superconductivity near 228 K, albeit with reduced volume fraction. Structural data at 136 GPa revealed minor shifts in Bragg peak positions of the $Fm\bar{3}m$ phase (Fig. S5), as previously discussed in the structural characterization section. These shifts likely reflect lattice relaxation and non-uniform pressure gradients, and correlate with the observed suppression of the higher-$T_c$ onset from 238 K at 142 GPa to 228 K at 136 GPa, a more rapid decline than typically reported in binary LaH$_{10}$ systems[9,10,12]. While pure LaH$_{10}$ transitions to lower-symmetry $C2/m$ or $R\bar{3}m$ phases near this pressure, no such transformations were evident here[6,12], suggesting that the observed $T_c$ suppression arises from phase dilution, lattice strain, and microstructural inhomogeneity, mechanisms known to influence superconductivity in clathrate hydrides[9,10,12,38].

## III. Conclusions

In this study, we demonstrated that partial yttrium substitution in LaH$_{10}$ enables the synthesis and stabilization of coexisting cubic $Fm\bar{3}m$ and hexagonal $P6_3/mmc$ clathrate phases in (La$_{0.9}$Y$_{0.1}$)H$_{10}$ across a wide pressure range, extending down to 136 GPa, well below the structural stability limit of pure LaH$_{10}$. Using synchrotron-based SXDM and XDI at the APS-U, we directly mapped μm-scale domain heterogeneity and visualized the spatial distribution of clathrate phases across the sample. By integrating these structural maps with multi-channel four-probe resistance measurements, we identified two distinct superconducting transitions: one near 244 K associated with $Fm\bar{3}m$-rich domains, and another near 220 K linked to $P6_3/mmc$-dominated regions. This spatially resolved correlation highlights how microscopic phase separation governs the macroscopic transport response in ternary superhydrides.

Our findings show that yttrium incorporation extends the structural stability of LaH$_{10}$-type clathrates without inducing secondary phases or substantially degrading the superconducting critical temperature. Furthermore, the combination of high-resolution structural imaging and spatially sensitive transport measurements provides a robust framework for probing electronic heterogeneity in high-pressure hydrides. More broadly, this work establishes a methodology for probing phase-separated superconductors with μm-scale resolution under extreme conditions. By bridging structural imaging and transport diagnostics, this approach opens new pathways for understanding emergent electronic phenomena in chemically substituted hydrides and informs design strategies aimed at optimizing superconductivity in multicomponent clathrate systems. An exciting next step would also be combining these techniques with new Miessner-effect imaging methods[39]. The successful stabilization and correlation of high-$T_c$ superconductivity in such complex materials mark a step toward the realization of practical hydrogen-based superconductors.

## IV. Methods

### A. Sample Preparation and High-Pressure Synthesis

A La–Y alloy with a nominal composition of La$_{0.9}$Y$_{0.1}$ was obtained from Ames Laboratory, prepared by arc melting high-purity lanthanum and yttrium metals under an argon atmosphere[40]. The stoichiometry and compositional homogeneity were verified using scanning electron microscopy and energy-dispersive X-ray spectroscopy (SEM–EDS) (Table S1, Fig. S10). Ambient-pressure X-ray diffraction confirmed the precursor metal was single phase (Fig. S11). DACs equipped with 65 μm culet diamonds were used for high-pressure synthesis and transport measurements. Sample loading was performed in an argon-filled glovebox to avoid air exposure. A piece of the La–Y alloy was placed in direct contact with ammonia borane (NH$_3$BH$_3$), which acted as both the hydrogen source and the pressure-transmitting medium[9,41]. To electrically isolate the sample from the metallic gasket, the sample chamber was prefilled with compressed cubic boron nitride (cBN) and laser drilled to create a sample chamber. Thin platinum foils (~2 μm thick) were manually positioned on the diamond culet to form electrical contacts in a VDP configuration for four-probe DC resistance measurements[30]. Secondary electrical connections were made externally using fine copper wires and silver epoxy. A schematic of the DAC assembly is provided in Fig. S2.

After sealing, the DACs were gradually compressed to the target synthesis pressures in multiple steps, with XRD measurements collected at intermediate pressures. Pressure was determined using the Raman shift of the diamond edge[42,43]. In situ laser heating was performed at 178 GPa on DAC #1 at beamline 13-ID-D of the APS, with results shown in Fig. S1. A second synthesis was conducted at 158 GPa using DAC #2 at beamline 16-ID-B of the APS-U; this sample was used for both XDI and transport measurements (Fig. 1). Laser heating in both cases was performed at the interface between the metal and ammonia borane using a modulated Yb fiber laser, delivering ~300 ms focused pulses that reached estimated temperatures of 1200–1800 K[44,45].

## B. Synchrotron X-ray Diffraction and Imaging

Synchrotron XRD measurements were performed at beamlines 13-ID-D of the APS and 16-ID-B of the APS-U at Argonne National Laboratory. Diffraction patterns were collected using PILATUS 2D area detectors and integrated into one-dimensional intensity profiles using DIOPTAS[46]. Structural parameters were extracted via Le Bail refinements using Jana2006[47,48]. This approach was necessary due to spotty diffraction and overlapping contributions from multiple coexisting phases, some with partially known or unknown structural models.

At 13-ID-D, the pre-upgrade X-ray beam was focused to ~3.8 × 2.7 μm². At 16-ID-B, the beam profile improved significantly with the APS-U upgrade: the beam size was reduced from ~6.3 × 3.1 μm² (pre-upgrade) to ~1.4 × 1.2 μm² (post-upgrade), with enhanced brilliance and minimal tails. This improvement enabled high-resolution spatial mapping with minimal signal overlapping between adjacent positions.

To visualize spatial phase distributions and structural inhomogeneity, SXDM was performed at 16-ID-B by raster scanning the laser-heated region over a two-dimensional grid. Each scan consisted of an 11 × 11 array of diffraction patterns acquired with step sizes of 2–5 μm. The collected datasets were processed using the X-ray Diffraction Imaging (XDI) software[26]. For each scanned point,

XDI converts the two-dimensional diffraction pattern into a radial intensity profile and integrates the signal within user-defined 2θ regions of interest corresponding to specific Bragg reflections. The resulting integrated intensities were compiled into two-dimensional maps, where each pixel represents a scanned position. The color intensity in these maps reflects the relative strength of the selected reflection, providing a direct visualization of local domain structure, spatial phase distribution, and structural heterogeneity[26,36].

## C. Electrical Transport Measurements

After synthesis and XRD analysis, the DAC was transferred to a continuous-flow cryostat for temperature-dependent resistance measurements down to 80 K. Data were collected during both cooling and warming cycles; however, only the warming cycle (at ~1 K/min) was analyzed due to reduced thermal gradients upon the slow warmup reducing temperature uncertainty. Resistance measurements were performed using a Keithley 6220 current source (100 µA excitation), a 2182A nanovoltmeter, and a Keithley 7001 switching matrix to acquire voltages across multiple electrode configurations[49]. Current and voltage leads were permuted in the standard VDP configuration[30]. Measured partial resistances are then defined as $R_{ab,cd} = \frac{V_{cd}}{I_{ab}}$, where $I_{ab}$ is the applied current between contacts a and b, and $V_{cd}$ being the voltage measured across contacts c and d. Because the sample did not satisfy the assumptions required for extracting sheet resistivity ($R_s$) using the standard VDP equation,

$$e^{\frac{-\pi(R_{12,34}+R_{34,12})}{2R_s}} + e^{\frac{-\pi(R_{23,41}+R_{41,23})}{2R_s}} = 1,$$

primarily due to sample inhomogeneity, the resultant $R_s$ values are not expected to represent the bulk electronic properties of a single phase. Nonetheless, VDP-averaged resistances are reported, as the superconducting transition is clearly observed (Fig. 3A). Correlations between individual partial resistances and the spatial phase distribution are explored in this work.

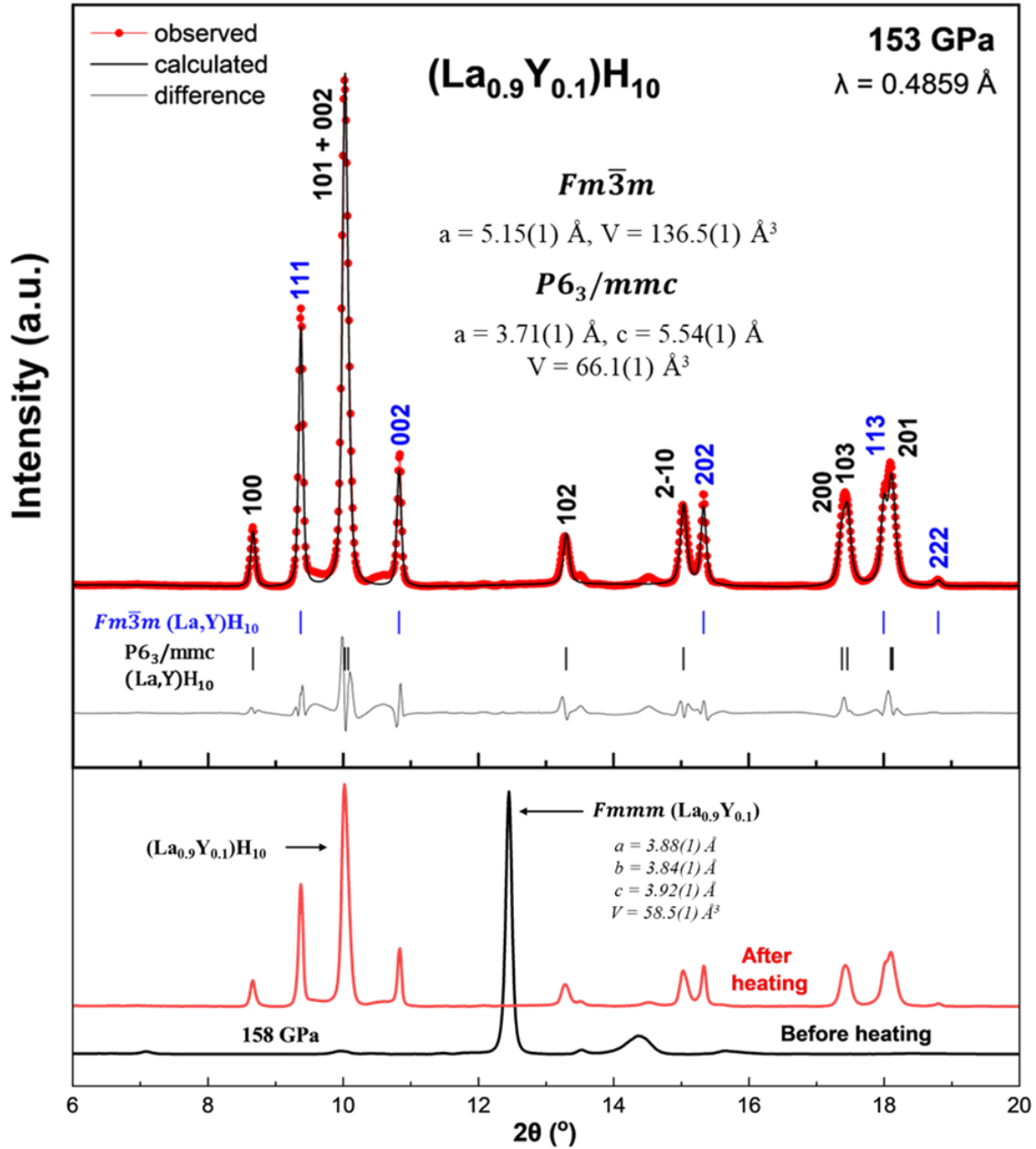

**Figure 1**: XRD patterns and structural refinement of coexisting (La,Y)H$_{10}$ phases. Top: Experimental synchrotron XRD patterns and Le Bail refinements of the $Fm\bar{3}m$ and $P6_3/mmc$ (La,Y)H$_{10}$ phases at 153 GPa. The experimental data, fit, and residuals are shown in red, black, and gray, respectively. Refined lattice parameters for both phases are indicated. Peaks corresponding to the Pt electrodes were excluded from the refinement. Bottom: Experimental XRD patterns of La$_{0.9}$Y$_{0.1}$ at 158 GPa before and after laser heating. The pre-heating pattern corresponds to the distorted-cubic Fmmm phase. After laser heating, the pressure decreased to 153 GPa, and the diffraction pattern shows the formation of (La,Y)H$_{10}$ phases, consistent with the refined structures shown above.

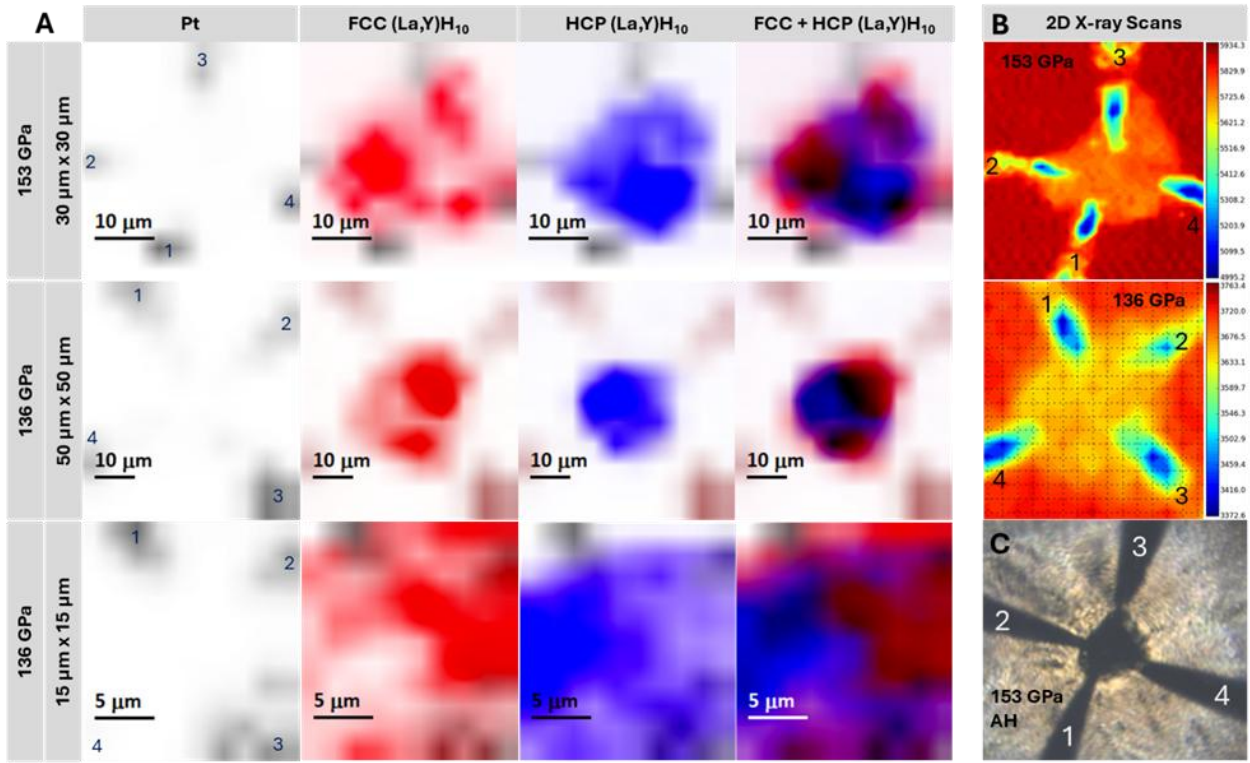

*Figure 2:* Spatially resolved XRD imaging of phase domains in $(La_{0.9}Y_{0.1})H_{10}$ at high pressure. (A) XDI maps of the laser-heated $(La,Y)H_{10}$ sample at 153 GPa (top) and 136 GPa (middle and bottom), obtained via raster scanning with a ~1 μm micro-focused synchrotron beam. The 30 μm × 30 μm scan at 153 GPa resolves spatial distributions of cubic $Fm\bar{3}m$ (red) and hexagonal $P6_3/mmc$ (blue) domains, with platinum leads mapped in dark brown. Electrode positions are annotated directly on the Pt phase map to enable spatial correlation with electrical transport measurements. The rightmost column overlays all phases to show the composite spatial distribution. At 136 GPa, a 50 μm × 50 μm scan (middle) and a higher-resolution 15 μm × 15 μm scan (bottom) show continued coexistence of FCC and HCP domains with spatial variation. (B) 2D X-ray scan overview showing the raster grid layout at respective pressures. (C) Optical image of the sample at 153 GPa after laser heating.

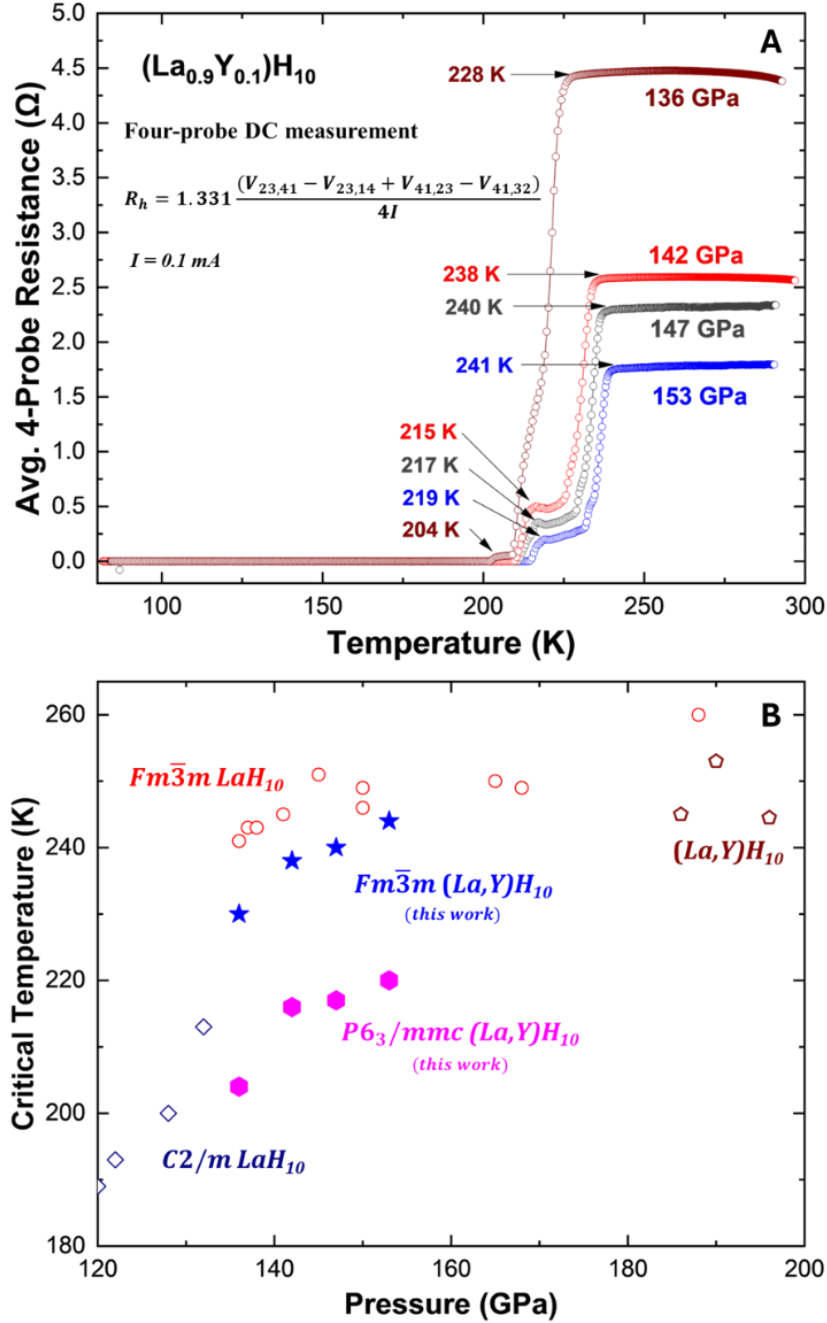

*Figure 3:* Superconducting behavior of (La,Y)$H_{10}$ during decompression and comparison with literature. (A) Temperature-dependent four-probe DC resistance measurements of (La,Y)$H_{10}$ collected at multiple pressures during decompression from 153 GPa to 136 GPa, using an excitation current of 0.1 mA. The curves represent the average four-probe resistance, calculated as shown in the inset to eliminate thermoelectric offsets. The pre-factor in the expression accounts for geometric correction in the VDP configuration. Distinct drops in resistance indicate superconducting transitions that persist across the entire pressure range. (B) Critical temperature ($T_c$) as a function of pressure for $LaH_{10}$ and (La,Y)$H_{10}$ hydrides. Open symbols represent $T_c$ values of $LaH_{10}$ and (La,Y)$H_{10}$ reported in literature[9–12,22]. Solid symbols indicate $T_c$ values of (La,Y)$H_{10}$ obtained in this work.

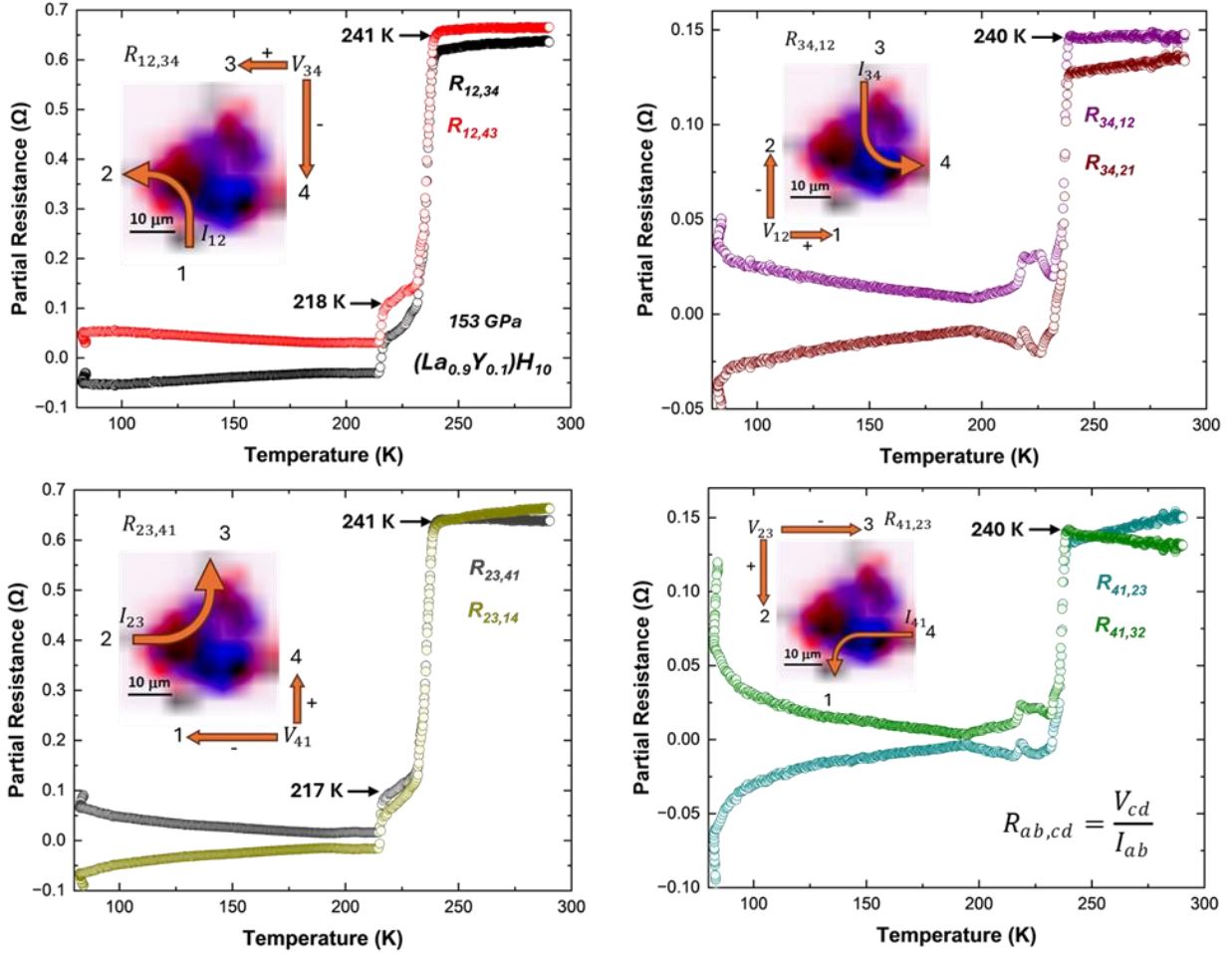

*Figure 4:* Spatial correlation between superconducting transitions and structural domains in $(La_{0.9}Y_{0.1})H_{10}$ at 153 GPa. Temperature-dependent four-probe partial resistance traces ($R_{ab,cd}$) measured using eight current–voltage configurations are shown across four panels. Each plot includes a schematic of the corresponding current path overlaid on the composite XDI map, highlighting the spatial distribution of cubic $Fm\bar{3}m$ (red) and hexagonal $P6_3/mmc$ (blue) phases. Electrode positions and current directions are annotated. Configurations that differ only by voltage polarity (e.g., $R_{34,12}$ vs. $R_{34,21}$) are spatially equivalent; for clarity, only one representative XDI map is shown for each pair. Partial resistance traces corresponding to current paths intersecting FCC-enriched domains (e.g., $R_{34,12}$, $R_{41,23}$) exhibit sharp superconducting transitions near 240 K with narrow widths ($\Delta T < 10$ K). In contrast, broader or multi-step transitions are observed in configurations that sample mixed-phase or HCP-dominated regions (e.g., $R_{12,34}$, $R_{23,14}$), with onsets near 241 K and 218 K. We note that due to single permutations inability to account for voltage drops due to the Seebeck effect, partial resistances exhibit negative values below $T_c$[49]. Averaging reverse-biased polarities corrects for this instrumental artifact, causing an apparent drop to zero resistance. These observations establish a direct spatial correlation between local structural heterogeneity and superconducting behavior.

**Data availability**
All relevant data are available from the corresponding authors upon request.

**References**

1. Boebinger, G. S. *et al.* Hydride superconductivity is here to stay. *Nat. Rev. Phys.* **7**, 2–3 (2025).
2. Ashcroft, N. W. Metallic hydrogen: a high-temperature superconductor? *Phys. Rev. Lett.* **21**, 1748–1749 (1968).
3. Ashcroft, N. W. Hydrogen dominant metallic alloys: high temperature superconductors? *Phys. Rev. Lett.* **92**, 187002 (2004).
4. Liu, H., Naumov, I. I., Hoffmann, R., Ashcroft, N. W. & Hemley, R. J. Potential high-$T_c$ superconducting lanthanum and yttrium hydrides at high pressure. *Proc. Natl. Acad. Sci. U.S.A.* **114**, 6990–6995 (2017).
5. Peng, F. *et al.* Hydrogen clathrate structures in rare earth hydrides at high pressures: possible route to room-temperature superconductivity. *Phys. Rev. Lett.* **119**, 107001 (2017).
6. Geballe, Z. M. *et al.* Synthesis and stability of lanthanum superhydrides. *Angew. Chem. Int. Ed.* **57**, 688–692 (2018).
7. Liu, H. *et al.* Dynamics and superconductivity in compressed lanthanum superhydride. *Phys. Rev. B* **98**, 100102 (2018).
8. Hemley, R. J., Ahart, M., Liu. H. & Somayazulu, M. Road to room-temperature superconductivity: $T_c$ above 260 K in lanthanum superhydride under pressure. In *Proc. Ramón Areces Symp. "Superconductivity and Pressure: A Fruitful Relationship on the Road to Room Temperature Superconductivity (Madrid, Spain, May 21–22, 2018).* (ed. Alario y Franco, M. A.) (2018).
9. Somayazulu, M. *et al.* Evidence for superconductivity above 260 K in lanthanum superhydride at megabar pressures. *Phys. Rev. Lett.* **122**, 027001 (2019).
10. Drozdov, A. P. *et al.* Superconductivity at 250 K in lanthanum hydride under high pressures. *Nature* **569**, 528–531 (2019).
11. Hong, F. *et al.* Superconductivity of lanthanum superhydride investigated using the standard four-probe configuration under high pressures. *Chin. Phys. Lett.* **37**, 107401 (2020).
12. Sun, D. *et al.* High-temperature superconductivity on the verge of a structural instability in lanthanum superhydride. *Nat. Commun.* **12**, 6863 (2021).
13. Wang, T. *et al.* Optimal alloying in hydrides: reaching room-temperature superconductivity in LaH$_{10}$. *Phys. Rev. B* **105**, 174516 (2022).
14. Hilleke, K. P. & Zurek, E. Rational design of superconducting metal hydrides via chemical pressure tuning. *Angew. Chem. Int. Ed.* **61**, e202207589 (2022).
15. Kostrzewa, M., Szczęśniak, K. M., Durajski, A. P. & Szczęśniak, R. From LaH$_{10}$ to room–temperature superconductors. *Sci. Rep.* **10**, 1–8 (2020).
16. Di Cataldo, S., von der Linden, W. & Boeri, L. First-principles search of hot superconductivity in La–X–H ternary hydrides. *npj Comput. Mater.* **8**, 1–8 (2022).
17. Denchfield, A., Park, H. & Hemley, R. J. Designing multicomponent hydrides with potential high $T_c$ superconductivity. *Proc. Natl. Acad. Sci. U. S. A.* **121**, e2413096121 (2024).

18. Sun, Y., Lv, J., Xie, Y., Liu, H. & Ma, Y. Route to a superconducting phase above room temperature in electron-doped hydride compounds under high pressure. *Phys. Rev. Lett.* **123**, 097001 (2019).
19. Di Cataldo, S., Heil, C., von der Linden, W. & Boeri, L. LaBH$_8$: Towards high-$T_c$ low-pressure superconductivity in ternary superhydrides. *Phys. Rev. B* **104**, L020511 (2021).
20. Liang, X. *et al.* Prediction of high-$T_c$ superconductivity in ternary lanthanum borohydrides. *Phys. Rev. B* **104**, 134501 (2021).
21. Zhao, W. *et al.* Superconducting ternary hydrides: progress and challenges. *Natl. Sci. Rev.* **10**, nwad307 (2023).
22. Semenok, D. V. *et al.* Superconductivity at 253 K in lanthanum–yttrium ternary hydrides. *Mater. Today* **48**, 18–28 (2021).
23. Bi, J. *et al.* Stabilization of superconductive La–Y alloy superhydride with $T_c$ above 90 K at megabar pressure. *Mater. Today Phys.* **28**, 100840 (2022).
24. Shen, G. & Mao, H. K. High-pressure studies with x-rays using diamond anvil cells. *Rep. Prog. Phys.* **80**, 016101 (2016).
25. Hemley, R. J., Mao, H. K. & Struzhkin, V. V. Synchrotron radiation and high pressure: new light on materials under extreme conditions. *J. Synchrotron Radiat.* **12**, 135–154 (2005).
26. Hrubiak, R., Smith, J. S. & Shen, G. Multimode scanning X-ray diffraction microscopy for diamond anvil cell experiments. *Rev. Sci. Instrum.* **90**, 025109 (2019).
27. Marshall, M. C. *et al.* Scanning X-ray diffraction microscopy for diamond quantum sensing. *Phys. Rev. Appl.* **16**, 054032 (2021).
28. Luo, A., Zhou, T., Holt, M. V., Singer, A. & Cherukara, M. J. Deep learning of structural morphology imaged by scanning X-ray diffraction microscopy. *Sci. Rep.* **15**, 21736 (2025).
29. Holt, M., Harder, R., Winarski, R. & Rose, V. Nanoscale hard X-ray microscopy methods for materials studies*. *Annu. Rev. Mater. Res.* **43**, 183–211 (2013).
30. van der Pauw, L. J. A method of measuring the resistivity and Hall coefficient on lamellae of arbitrary shape. *Philips Tech. Rev.* **20**, 220–224 (1958).
31. van der Pauw, L. J. A method of measuring specific resistivity and Hall effect of discs of arbitrary shapes. *Philips Res. Rep.* **13**, 1–9 (1958).
32. Chen, W. *et al.* Superconductivity and equation of state of lanthanum at megabar pressures. *Phys. Rev. B* **102**, 134510 (2020).
33. Gavriliuk, A. G. *et al.* Synthesis and magnetic properties of iron polyhydrides at megabar pressures. *JETP Lett.* **116**, 804–816 (2022).
34. Du, F. *et al.* Superconducting gap of H$_3$S measured by tunnelling spectroscopy. *Nature* 1–6 (2025).
35. Ji, C. *et al.* Ultrahigh-pressure isostructural electronic transitions in hydrogen. *Nature* **573**, 558–562 (2019).
36. Chen, S. *et al.* Superior superconducting properties realized in quaternary La–Y–Ce hydrides at moderate pressures. *J. Am. Chem. Soc.* **146**, 14105-14113 (2024).

37. Deemyad, S. *et al.* Dependence of the superconducting transition temperature of single and polycrystalline MgB$_2$ on hydrostatic pressure. *Physica C* **385**, 105–116 (2003).
38. Troyan, I. A. *et al.* Anomalous high-temperature superconductivity in YH$_6$. *Adv. Mater.* **33**, 2006832 (2021).
39. Bhattacharyya, P. *et al.* Imaging the meissner effect in hydride superconductors using quantum sensors. *Nature* **627**, 73–79 (2024).
40. Gschneidner, K. A. & Calderwood, F. W. The La−Y (lanthanum−yttrium) system. *Bull. Alloy Phase Diagr.* **3**, 94–96 (1982).
41. Staubitz, A., Robertson, A. P. M. & Manners, I. Ammonia-borane and related compounds as dihydrogen sources. *Chem. Rev.* **110**, 4079–4124 (2010).
42. Hanfland, M. & Syassen, K. A Raman study of diamond anvils under stress. *J. Appl. Phys.* **57**, 2752–2756 (1985).
43. Akahama, Y. & Kawamura, H. Pressure calibration of diamond anvil Raman gauge to 410 GPa. *J. Phys.: Conf. Ser.* **215**, 012195 (2010).
44. Meng, Y., Hrubiak, R., Rod, E., Boehler, R. & Shen, G. New developments in laser-heated diamond anvil cell with in situ synchrotron X-ray diffraction at High Pressure Collaborative Access Team. *Rev. Sci. Instrum.* **86**, 072201 (2015).
45. Prakapenka, V. B. *et al.* Advanced flat top laser heating system for high pressure research at GSECARS: application to the melting behavior of germanium. *High Press. Res.* **28**, 225–235 (2008).
46. Prescher, C. & Prakapenka, V. B. DIOPTAS: a program for reduction of two-dimensional X-ray diffraction data and data exploration. *High Press. Res.* **35**, 223–230 (2015).
47. Le Bail, A. Whole powder pattern decomposition methods and applications: a retrospection. *Powder Diffr.* **20**, 316–326 (2005).
48. Petříček, V., Dušek, M. & Palatinus, L. Crystallographic computing system JANA2006: general features. *Z. Kristallogr. Cryst. Mater.* **229**, 345–352 (2014).
49. Keithley, J. F. & Keithley Instruments Inc. *Low Level Measurements Handbook: Precision DC Current, Voltage, and Resistance Measurements* (Keithley Instruments, 2004).

**Acknowledgements**

This work was supported by the US Department of Energy (DOE)-National Nuclear Security Administration (NNSA; DE-NA0004153, Chicago/DOE Alliance Center), National Science Foundation (NSF, DMR-2118020), and DOE-Office of Science (DE-SC0020340). Operations of HPCAT (Sector 16, APS, ANL) are supported by DOE-NNSA's Office of Experimental Sciences. Operations of GeoSoilEnviroCARS (The University of Chicago, Sector 13, APS, ANL) are supported by the National Science Foundation – Earth Sciences via SEES: Synchrotron Earth and Environmental Science (EAR –2223273). Work at the Advanced Photon Source was supported by

**Author contributions**

A.H.M.M. and R.J.H. designed the research. A.H.M.M., K.W., N.P.S., M.H. prepared the experiments. A.H.M.M., K.W., N.P.S., M.H., R.H., S.C., D.S., V.B.P., M.S., N.V. performed the synchrotron-based experiments. A.H.M.M., A.C.M., K.W., N.P.S., M.H., R.J.H. conducted the four- probe electrical transport measurements. A.H.M.M., K.W., N.P.S, A.C.M., M.H., R.J.H. analyzed data. All authors participated in discussing the results and writing the paper.

**Competing interests**

The authors declare no competing interests.

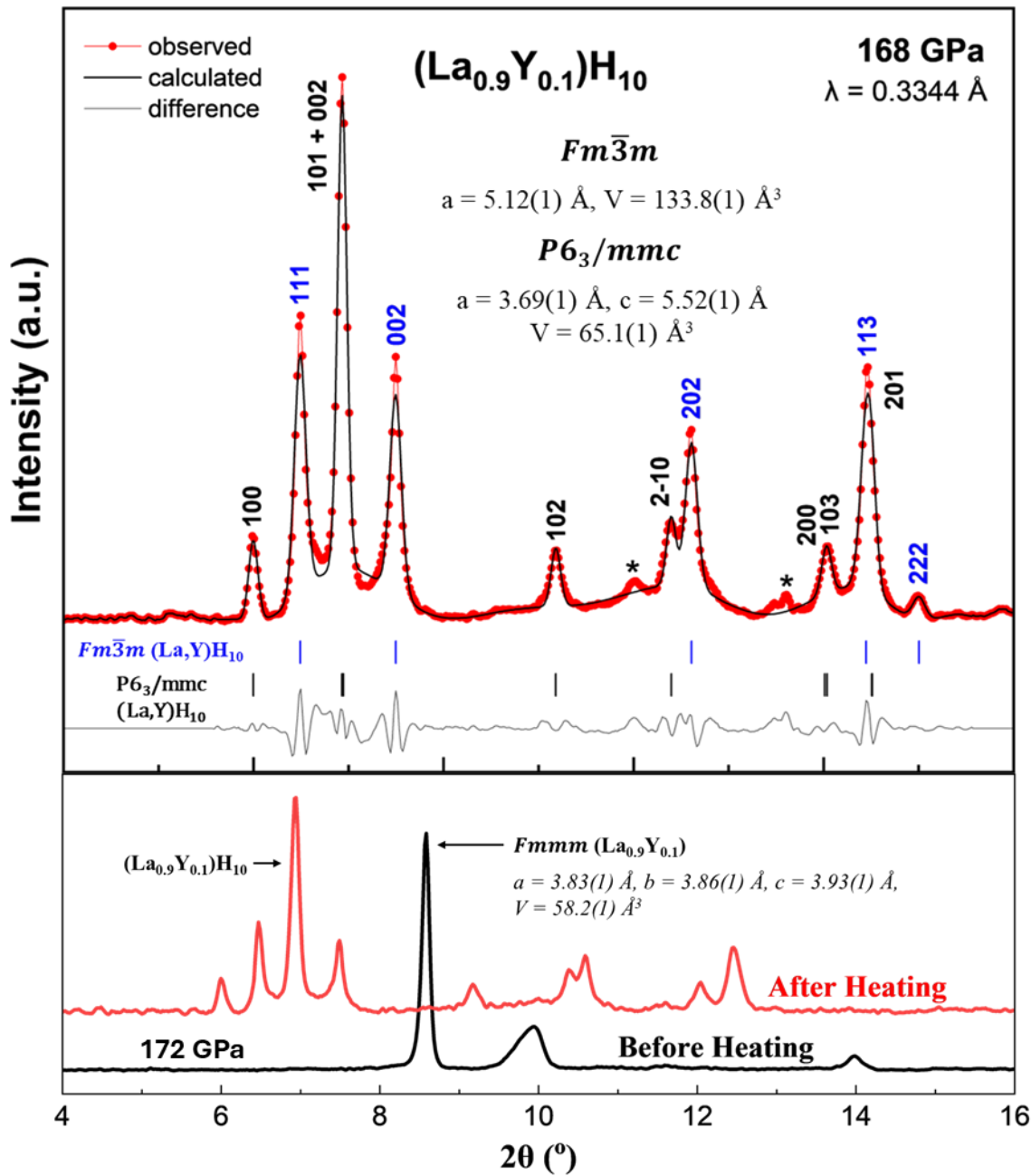

*Figure S1:* XRD patterns and structural refinement of coexisting (La,Y)$H_{10}$ phases at 168 GPa. Top: Experimental synchrotron X-ray diffraction patterns and Le Bail refinements of the $Fm\bar{3}m$ and $P6_3/mmc$ (La,Y)$H_{10}$ phases at 168 GPa. The experimental data, fit, and residuals are shown in red, black, and gray, respectively. Refined lattice parameters for both phases are indicated. Bottom: Experimental X-ray diffraction patterns of La$_{0.9}$Y$_{0.1}$ at 172 GPa before and after laser heating. The pre-heating pattern corresponds to the distorted FCC phase. The post-heating pattern shows the formation of (La,Y)$H_{10}$, with the top refinement corresponding to the (La,Y)$H_{10}$ phases.

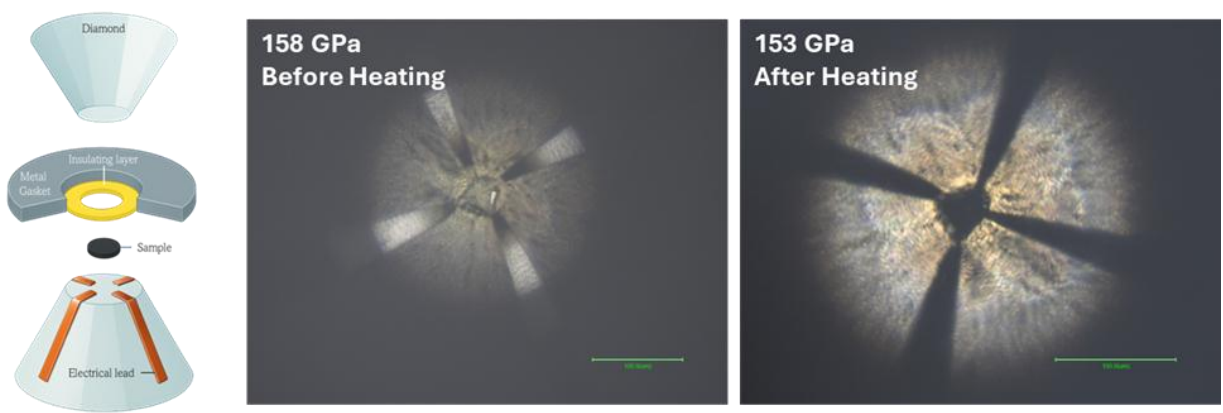

***Figure S2:*** *(Left) Schematic illustration of the diamond anvil cell (DAC) assembly configured for four-probe electrical transport measurements. (Right) Optical images of the (La,Y)H$_{10}$ DAC #2 sample at 158 GPa before laser heating (transmitted and reflected light) and after heating at 153 GPa (transmitted light). A clear volume expansion is observed following synthesis, as evidenced by the transition of the initially transparent NH$_3$BH$_3$ region to an opaque state on the right side of the culet.*

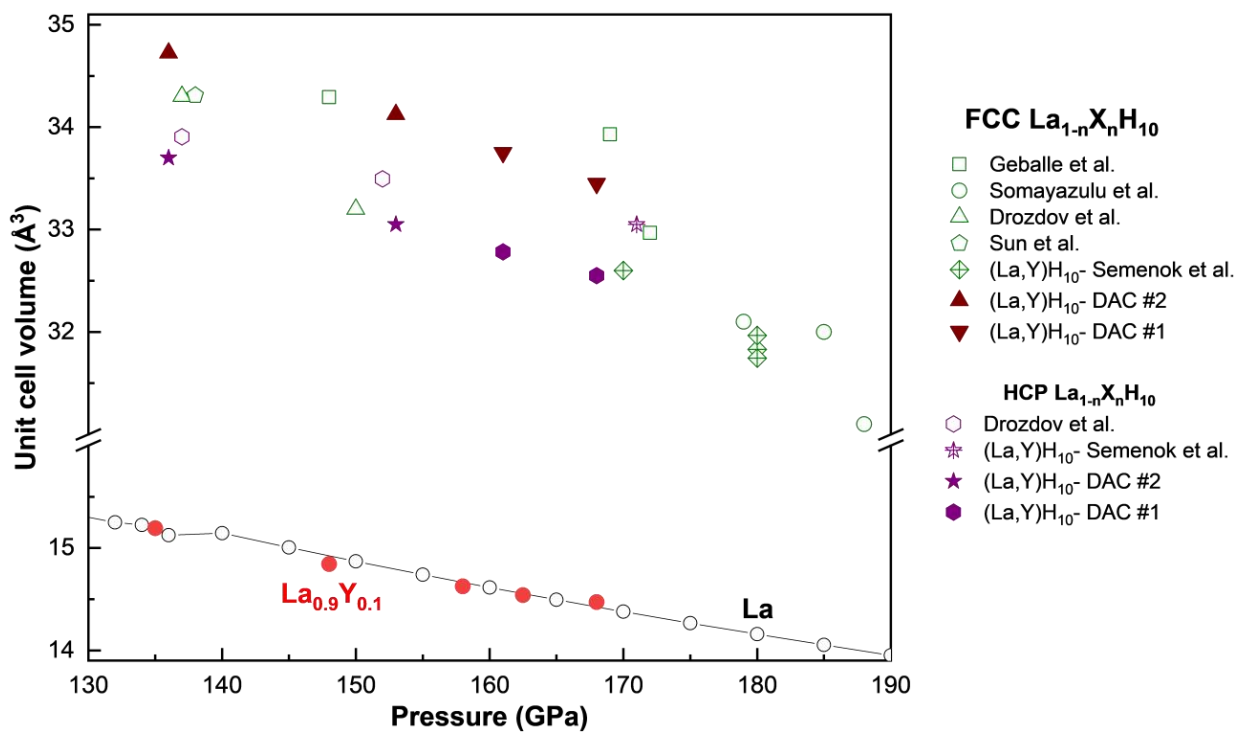

***Figure S3:*** *Pressure–unit cell volume comparison of La, La$_{0.9}$Y$_{0.1}$, LaH$_{10}$, and (La,Y)H$_{10}$ phases. The figure shows experimental data collected in this study and from previous literature[6,9,10,12,22]. Filled symbols represent data from this work, while open symbols correspond to literature values. Pressure calibration for our measurements was performed using the diamond Raman edge method. Some literature data were calibrated using the hydrogen vibron[10,12], which may underestimate the true pressure, leading to actual pressures being higher than reported.*

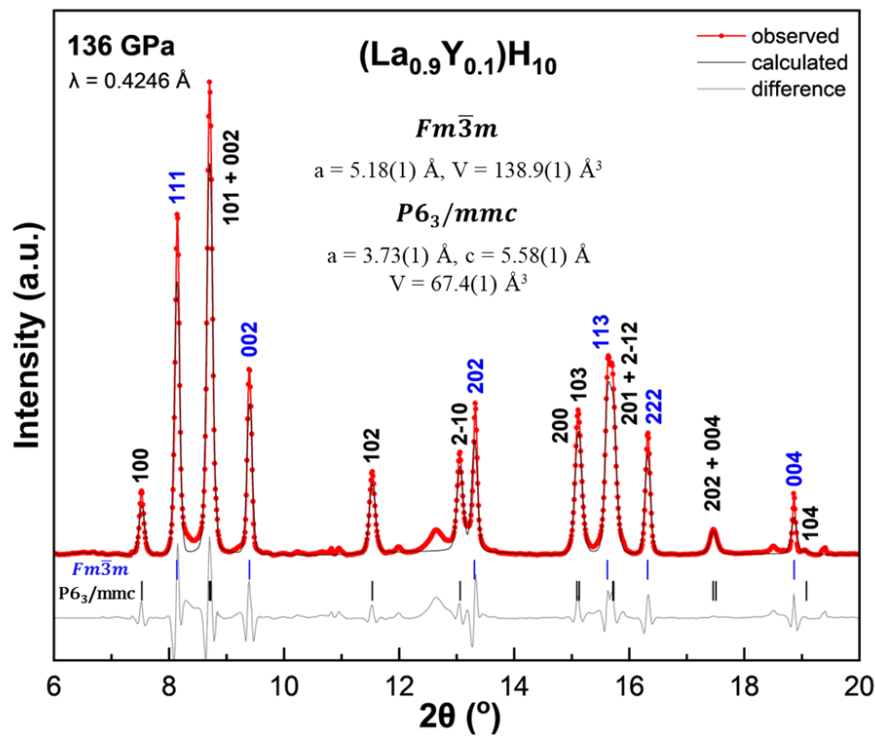

*Figure S4:* Experimental synchrotron X-ray diffraction patterns and Le Bail refinements of the $Fm\bar{3}m$ and $P6_3/mmc$ (La,Y)H$_{10}$ phases at 136 GPa. The experimental data, fit, and residuals are shown in red, black, and gray, respectively. Refined lattice parameters for both phases are indicated.

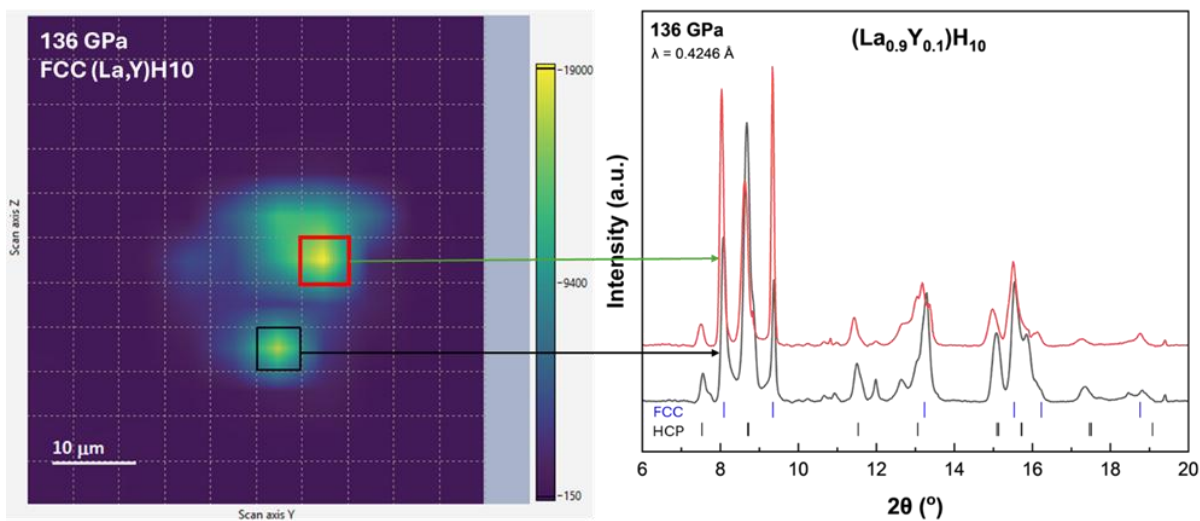

*Figure S5:* Spatially resolved X-ray diffraction analysis of the cubic phase at 136 GPa. X-ray diffraction imaging (XDI) map of the $Fm\bar{3}m$ (cubic) phase of (La,Y)H$_{10}$ at 136 GPa, showing its spatial distribution across the scanned region. Diffraction patterns extracted from two high-intensity regions of the grid (right) confirm the presence of the cubic phase with slight peak shifts between locations, likely arising from local pressure gradients and lattice relaxation effects.

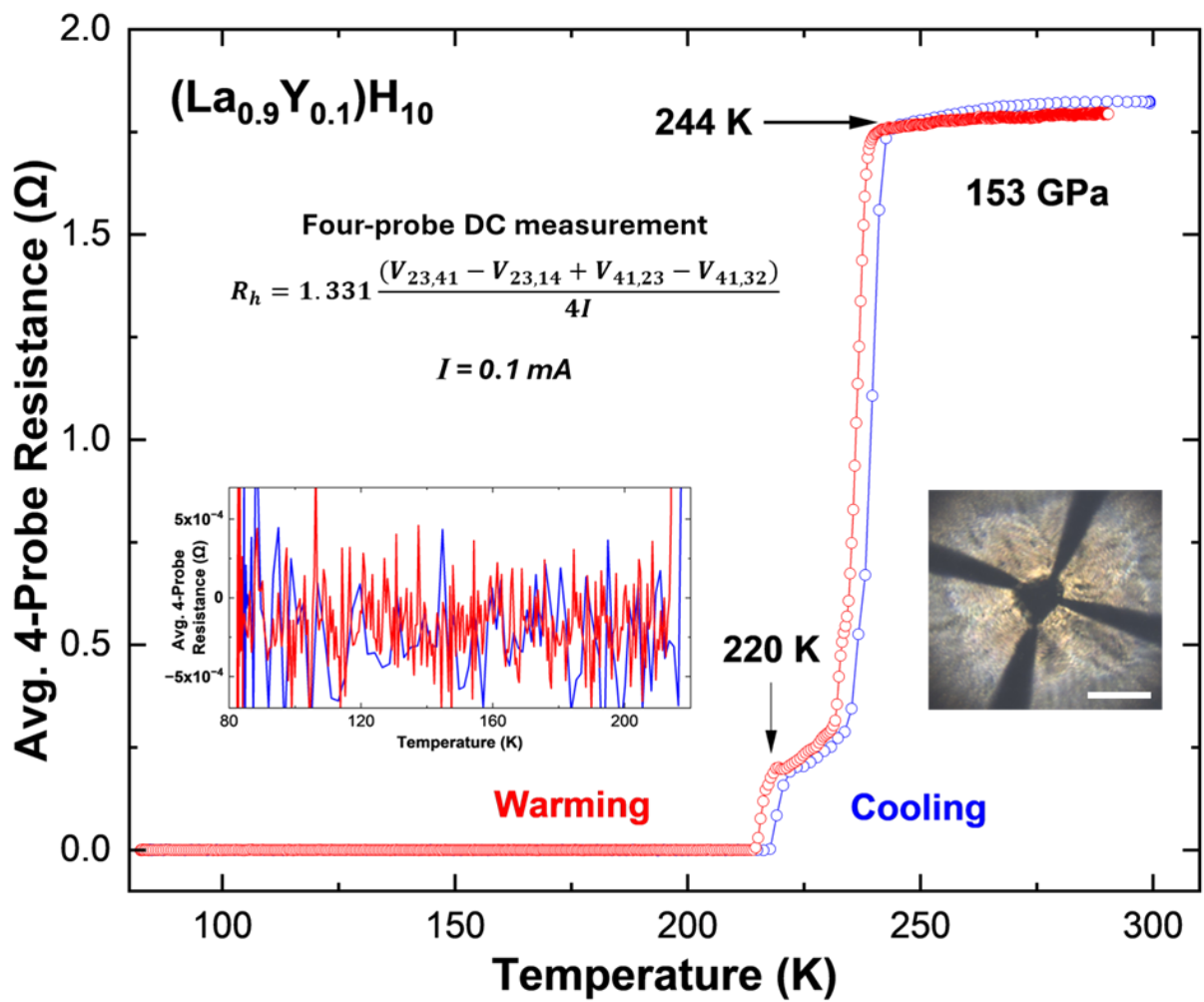

*Figure S6:* R–T of (La,Y)H$_{10}$ at 153 GPa. Temperature-dependent resistance of (La$_{0.9}$Y$_{0.1}$)H$_{10}$ measured using a four-probe DC configuration at 153 GPa with an excitation current of 0.1 mA. The plot shows both cooling and warming cycles, revealing two distinct superconducting transitions at approximately 244 K and 220 K. The curves represent the average four-probe resistance, calculated using the expression shown in the inset to eliminate thermoelectric offsets. The pre-factor accounts for geometric correction in the VDP configuration. Insets show a magnified view demonstrating that all signals drop below the instrument's noise floor in the superconducting state, and an optical image of the DAC culet with platinum electrodes.

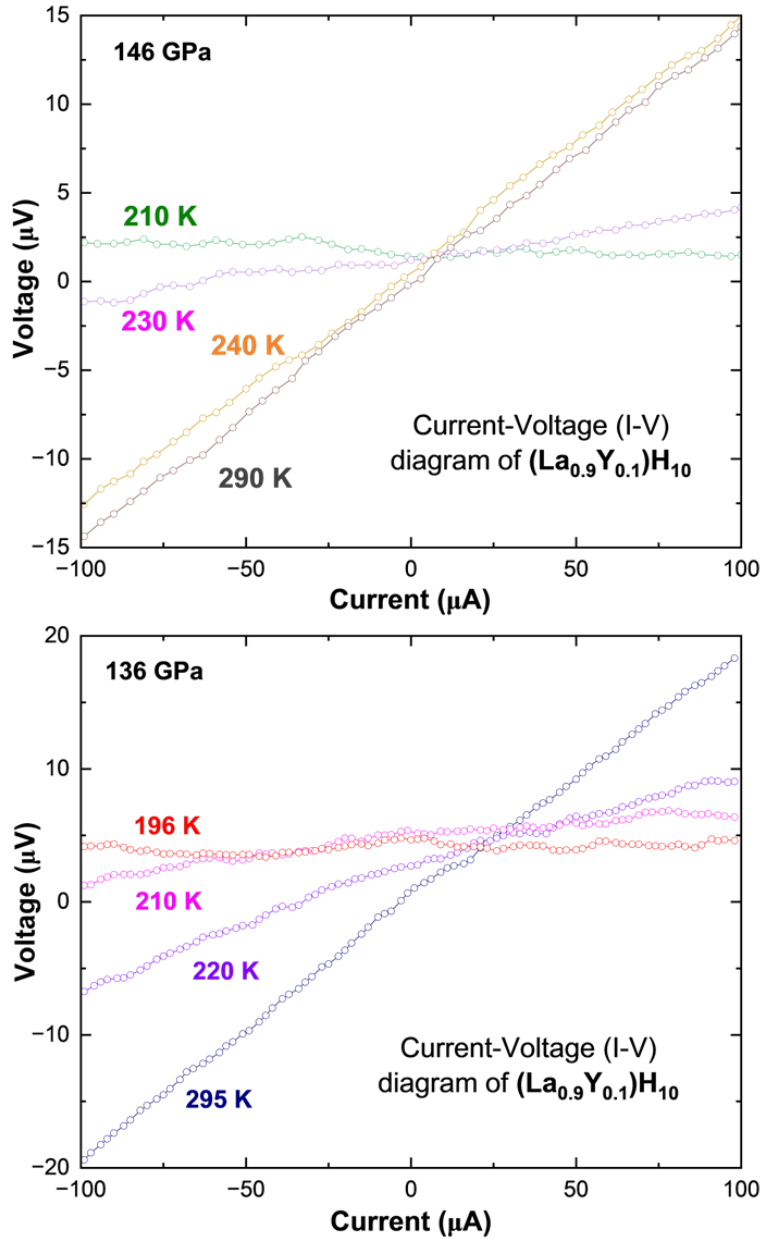

*Figure S7:* Current–voltage (I–V) characteristics of $(La_{0.9}Y_{0.1})H_{10}$ measured at 146 GPa and 136 GPa across various temperatures, using a delta-mode Keithley 6220 current source and 2182A nanovoltmeter[49]. At 146 GPa, the I–V response is linear at 290 K, consistent with metallic Ohmic behavior. Below ~230 K, the curves become increasingly nonlinear, with a pronounced deviation at 210 K, mirroring the superconducting transition seen in resistance–temperature data. At 136 GPa, similar non-Ohmic behavior emerges below 210 K, indicating persistent superconductivity upon decompression. The nonlinear I–V profiles reflect the formation of a superconducting gap and critical current limitations, consistent with prior high-pressure studies[9]. These measurements provide complementary confirmation of superconductivity in $(La,Y)H_{10}$ and highlight the utility of I–V analysis alongside resistance and structural mapping in mixed-phase superhydrides.

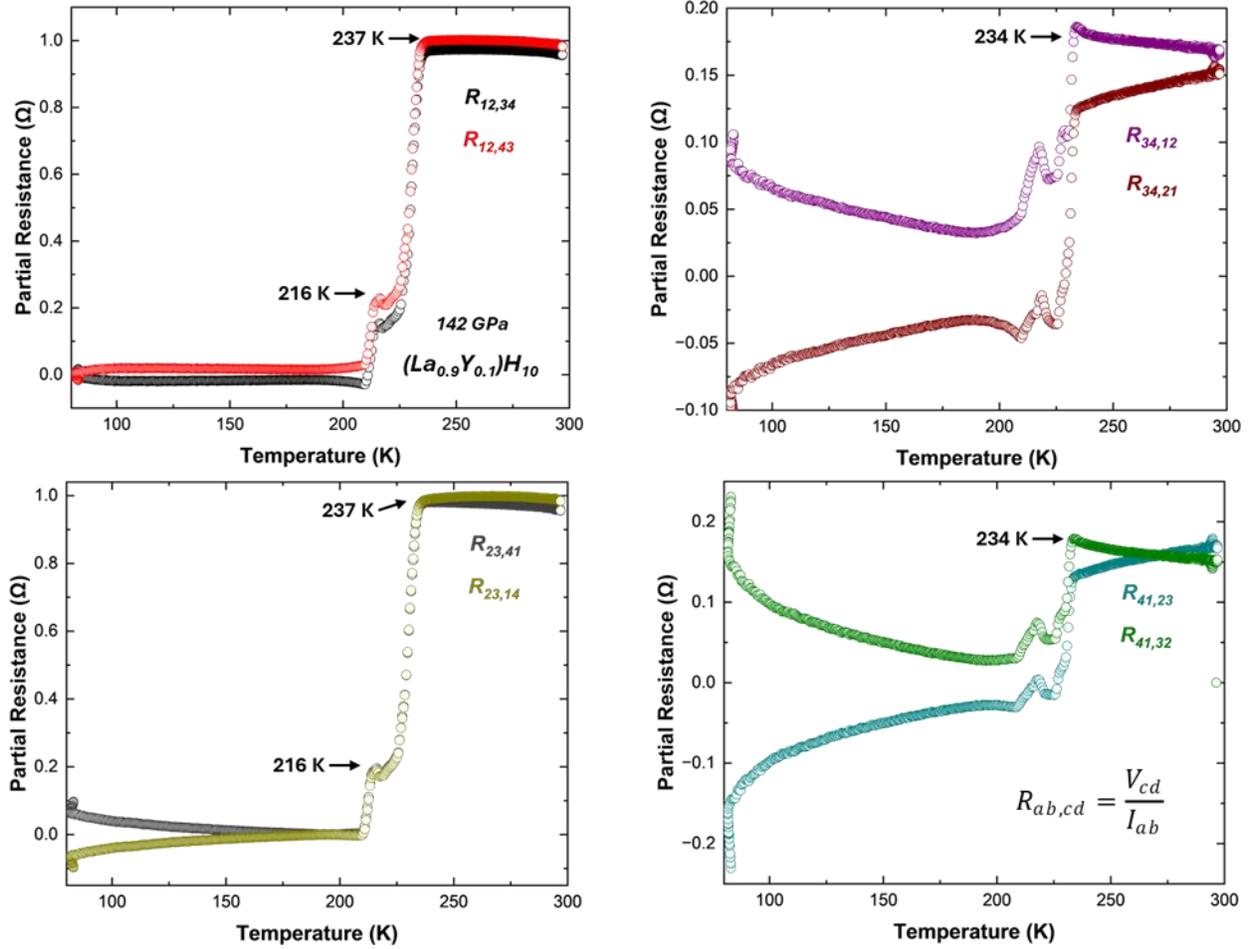

*Figure S8:* Temperature-dependent four-probe partial resistance traces ($R_{ab,cd}$) of $(La_{0.9}Y_{0.1})H_{10}$ at 142 GPa, measured using eight current–voltage configurations are shown across four panels. Traces corresponding to current paths intersecting FCC-enriched regions (e.g., $R_{34,12}$, $R_{41,23}$) exhibit sharp transitions near 240 K with $\Delta T < 10$ K, while configurations sampling mixed or HCP-rich regions (e.g., $R_{12,34}$, $R_{23,14}$) show broader or multi-step transitions. Negative resistance values below $T_c$ arise from thermoelectric (Seebeck) voltage offsets that dominate the signal after the sample becomes superconducting[49]. These results further support the link between local structural domains and superconducting behavior.

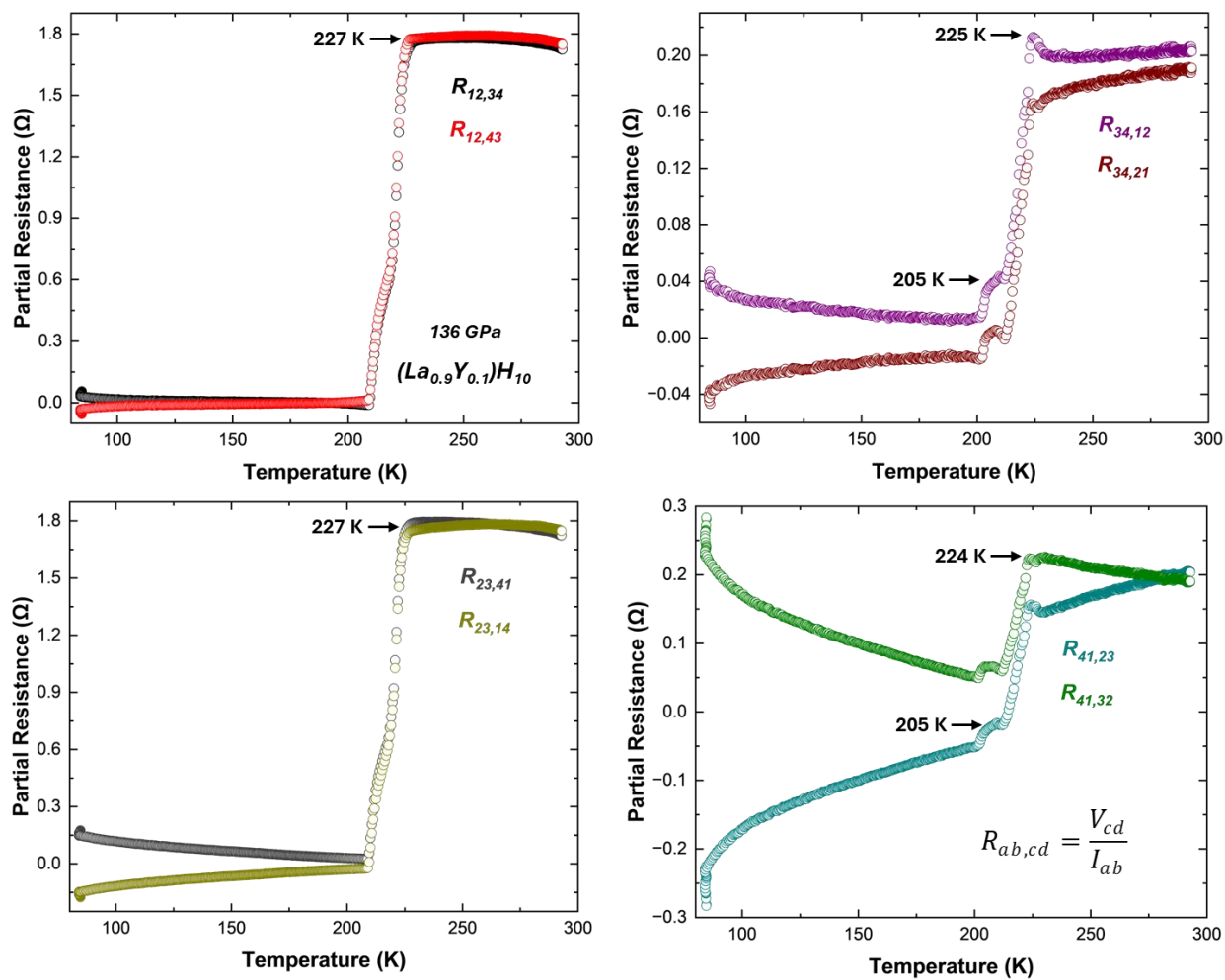

***Figure S9:*** *Temperature-dependent four-probe partial resistance traces ($R_{ab,cd}$) of $(La_{0.9}Y_{0.1})H_{10}$ at 136 GPa, measured using eight current–voltage configurations are shown across four panels.*

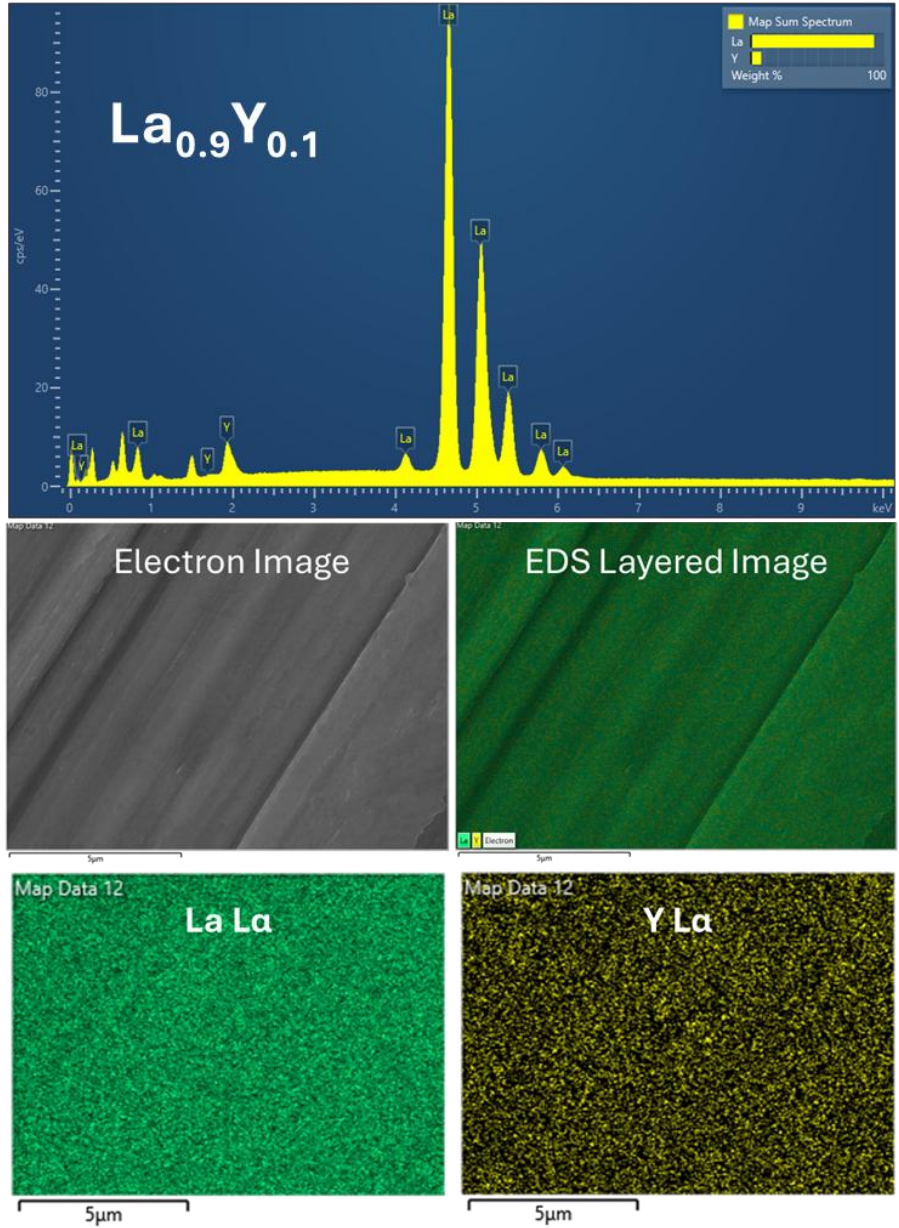

***Figure S10:*** *SEM-EDS characterization of the La$_{0.9}$Y$_{0.1}$ alloy. Top row: EDS spectrum acquired at 5 µm resolution, showing characteristic peaks of La (Mα at ~0.83 keV, Lα at ~4.65 keV, and Lβ at ~5.48 keV) and Y (Lα at ~1.92 keV). Middle row: Secondary electron image of the region where EDS data were collected (left), followed by layered EDS images highlighting the distribution of detected elements. Bottom row: Elemental maps showing the spatial distribution of La (La Lα) and Y (Y Lα), confirming a uniform dispersion of both elements across the analyzed region.*

*Table S1:* Results of SEM-EDS analysis for $La_{0.9}Y_{0.1}$ alloy. The table includes measured values for each element: line type, apparent concentration, k-ratio, weight percentage (Wt%), standard deviation (Wt% Sigma), and reference standard label.

| $La_{0.9}Y_{0.1}$ | | | | | | |
|---|---|---|---|---|---|---|
| Element | Line Type | Apparent Concentration | k-Ratio | Wt% | Wt% Sigma | Standard Label |
| Y | L series | 1.08 | 0.01082 | 7.67 | 0.15 | Y |
| La | L series | 35.12 | 0.31512 | 92.33 | 0.15 | $LaB_6$ |
| Total: | | | | 100.00 | | |

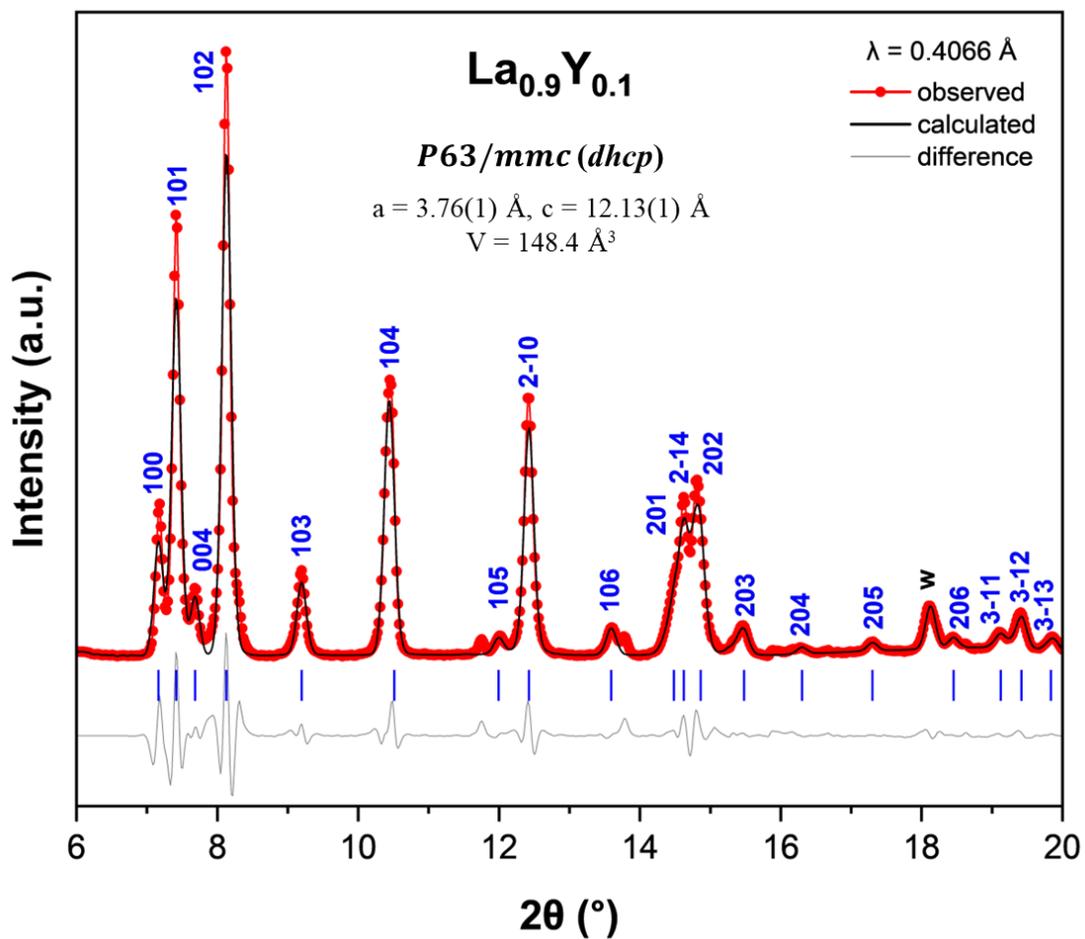

**Figure S11:** Synchrotron XRD patterns of La$_{0.9}$Y$_{0.1}$ alloy at ambient conditions. The La$_{0.9}$Y$_{0.1}$ sample exhibits a dhcp structure. The patterns confirm phase purity and show no evidence of phase separation or residual elemental La or Y, indicating successful alloy formation.